\documentclass[prl,twocolumn,amsmath,amssymb,nofootinbib]{revtex4-1}
\usepackage{amsmath}
\usepackage{amssymb}
\usepackage{amsfonts}
\usepackage{eucal}

\newcommand{\be}{\begin{equation}}
\newcommand{\bse}{\begin{subequations}}
\newcommand{\ese}{\end{subequations}}
\newcommand{\bea}{\begin{eqnarray}}
\newcommand{\eea}{\end{eqnarray}}
\newcommand{\ba}{\begin{array}}
\newcommand{\ea}{\end{array}}
\newcommand{\ee}{\end{equation}}

\def\Pl{{\rm Pl}}
\def\hp{h_+}
\def\hc{h_{\times}}
\def\half{\frac{1}{2}}

\def\yboxit#1#2{\vbox{\hrule height #1 \hbox{\vrule width #1
    \vbox{#2}\vrule width #1 }\hrule height #1 }}
\def\fillbox#1{\hbox to #1{\vbox to #1{\vfil}\hfil}}
\def\ybox{\yboxit{0.4pt}{\fillbox{8pt}}\hskip-0.4pt}
\def\VEV#1{\langle{ #1} \rangle}

\begin{document}

%\preprint{
%           hep-th/yymmnnn \cr
%           SLAC-PUB-xxxx\cr
%           SU-ITP-04/nn\cr
%}
%{\vskip .2cm}
%\vspace*{2mm}

\title{ Leptogenesis from Gravity Waves in Models of Inflation}

\author{Stephon H.S Alexander, Michael E. Peskin}
%\email{stephon, mpeskin@slac.stanford.edu}
\affiliation{Stanford Linear Accelerator Center,
P.O.Box 20450, Stanford,CA 94309 USA}
\author{M.M. Sheikh-Jabbari}
\affiliation{Department of Physics, Stanford University,
382 via Pueblo Mall, Stanford CA 94305-4060, USA}
%\email{jabbari@itp.stanford.edu}

\begin{abstract}
We present a new mechanism for creating the observed cosmic matter-antimatter
asymmetry  which satisfies all
three Sakharov conditions from one common thread, gravitational
waves.   We generate lepton number through the gravitational anomaly in
the lepton number current.  The source term comes from elliptically
polarized gravity waves that are produced during inflation if the
inflaton field contains a CP-odd component.
The
amount of matter asymmetry generated in our model can be of realistic size for
the parameters within the range of some inflationary scenarios and grand
unified theories.
%In simple inflationary scenarios, the generated matter asymmetry is very
%small.  We describe some
%special conditions in which our mechanism can give a matter
%asymmetry of realistic size.
\end{abstract}
%\pacs{98.80Cq, 98.80.Es, 11.30.Er}
%\keywords{Leptogenesis, Gravitational Waves, Anomaly}
%\date{\today}
%\date{October 2003}
\maketitle

%\section{Introduction}

As far into the universe as we can see, there is an excess of matter over
anti-matter.  The recent determinations of the cosmological parameters
from the cosmic microwave background by the WMAP experiment gives the
baryon density of the universe as \cite{WMAP}
\be\label{baryond}
%\frac{n_b}{n_\gamma} = (6.5\pm 0.4)\times 10^{-10}\ .
{n_b}/{n_\gamma} = (6.5\pm 0.4)\times 10^{-10}\ .
\ee
This is a small number, but at the same time it is large enough to be a
puzzle for models of particle physics.  A baryon excess this large cannot be
produced in the early universe within the Standard Model of particle physics
\cite{HSather}.
In this
paper, we introduce a new mechanism for the creation of the matter-antimatter
asymmetry, one associated with gravitational fluctuations created during
cosmological inflation.

The conditions for generating a matter-antimatter asymmetry were stated by
Sakharov almost forty years ago~\cite{Sakharov}.
 First, baryon number should be violated.
  Second, CP should be violated.  Third, these symmetry violations should be
relevant at a time when the universe is out of thermal equilibrium.
 Since the 1980's,
it has been realized that the standard weak interactions contain processes,
mediated by {\it sphaelerons}, which interconvert baryons and leptons and are
thermally activated at temperatures greater than 1 TeV.  Thus, we can also
create the baryon asymmetry by creating net lepton number at high temperature
through out-of-equilibrium and CP-asymmetric processes \cite{KRS,FY}.
  Scenarios of this
type are known as {\it leptogenesis}.

The out-of-equilibrium conditions can be created at a
phase transition or through
late decay of massive particles.  The most attractive
choice for a phase transition
is that associated with electroweak symmetry breaking.  However, that
phase transition is probably not sufficiently strongly first-order.
This is known
to be an obstacle to baryogenesis in the Standard Model, and most of the
allowed parameter space in minimal supersymmetry is also already
excluded.
Particle decay asymmetries are loop-suppressed and therefore require
relatively large CP-violating phases.  Such large phases are strongly
constrained in supersymmetry \cite{SUSYphases} though they still could appear
 in the neutrino Yukawa couplings that are used in the  Fukugita-Yanagida
scenario for
leptogenesis \cite{FY}.   In any event,
there is good reason to seek more effective
sources of CP-violating out-of-equilibrium physics.

%\vspace*{1mm}
%\begin{center}
%{\bf{Outline of the mechanism}}
%\end{center}
%\vspace*{1mm}
%\section{Outline of the mechanism}

Our model of matter-antimatter asymmetry is assembled out of the
following ingredients.
%We would like to assemble a theory of matter-antimatter asymmetry out of the
%following ingredients.
First, as is well-known \cite{AGW},
the lepton number current, and
also the total fermion number current, has a gravitational anomaly in the
Standard Model.  Explicitly,
\be\label{Jlepton}
     \partial_\mu J^\mu_\ell  =  \frac{3}{16\pi^2}   R  \tilde R
\ee
where
\be
J^\mu_\ell =    \bar \ell_i\gamma^\mu \ell_i + \bar \nu_i \gamma^\mu \nu_i \  ,\
R\tilde R = \half \epsilon^{\alpha\beta\gamma\delta}
 R_{\alpha\beta \rho\sigma} R_{\gamma\delta}{}^{\rho\sigma} \ .
\ee
The anomaly requires an imbalance of left- and right-handed leptons, so we
are ignoring right-handed neutrinos.  In general, \eqref{Jlepton} will be
correct in an effective theory valid below a scale $\mu$.  A simple guess
for $\mu$ is that it is at the right-handed neutrino scale,
of the order of $10^{14}$ GeV, but we would like to keep in mind
the possibility of higher values of $\mu$.

 Next, we
claim, a contribution to $R\tilde R$ of definite sign can be generated by
gravitational fluctuations produced during inflation if the inflaton field
contains a CP-odd component.  This can be naturally achieved if the inflaton is a
complex modulus field such as one finds in supergravity or superstring models.
Such fields can have the very flat potentials required for inflation.
The imaginary
part $\phi$ of this field (which we henceforth call an `axion') can couple
to gravity through an interaction
\be\label{axioncoupling}
     \Delta {\cal L} =    F(\phi)  R  \tilde R  \ ,
\ee
where $F$ is odd in $\phi$,
as a result of the Green-Schwarz mechanism \cite{GS}.  Lue, Wang, and
Kamionkowski (LWK) have studied the effects of such an interaction in
generating observable parity-violation in the cosmic microwave background
\cite{LWK}.   A simple form for
$F(\phi)$ is
%with the correct scaling is
\be\label{Fval}
%       F(\phi) =   {1\over 16\pi^2 M_\Pl} {\cal N} \phi \ ,
       F(\phi) ={\cal N} \phi /{(16\pi^2 M_\Pl)}\ ,
\ee
where ${\cal N}$ is the number of stringy degrees of freedom propagating in the loops and
the $M_\Pl$ in the denominator is approximately the string scale.
In principle, this $M_\Pl$ can be  substituted with a lower mass scale ${\cal F}$
with the constraint that our effective
field theory is valid only for $\mu < {\cal F}$.

We would like to apply the interaction  \eqref{axioncoupling}
to the dynamics of metric
fluctuations during inflation.
When the axion field
has a slowly-rolling nonzero classical value, the coupling
\eqref{axioncoupling}  can lead to quantum
fluctuations of the gravitational field that, treated to second order,
 generate a nonzero right-hand
side for \eqref{Jlepton}.

We believe that our
analysis is also interesting because
the Sakharov conditions are satisfied in this scenario in an unusual way.
Lepton number is violated through \eqref{Jlepton}.  CP violation
and out-of-equilibrium result from the nonzero classical value of the axion
field.  Before inflation, the complex modulus field varies from point to point
in both modulus and phase.  Inflation blows up a small region in this
field to a
size much greater than that of the visible universe.  In this region, the
modulus field is approximately constant and has a randomly chosen, fixed
phase.  This value then rolls slowly toward the minimum of its potential.
In this process, we have out-of-equilibrium dynamics and, if the phase is
nonzero, a CP asymmetry.  We claim that {\it no explicit CP violation is
needed in the equations of motion}.  The CP-odd field $\phi$ could have
zero expectation value today and need have no relation to the CP violation
observed in particle physics.

Now we would like to quantitatively estimate the lepton number produced in
inflation~\cite{LL}.
The general form of metric perturbations about an FRW universe can be
parameterized as
\be
\begin{split}
ds^2 &= -(1+2\varphi)dt^2+w_i dtdx^i\cr
&+a^2(t)\left[\left((1+2\psi)\delta_{ij}+h_{ij}\right)dx^idx^j\right]
\end{split}
\ee
where $\varphi$, $\psi$, $w_i$ and $h_{ij}$ respectively parameterize the
 scalar, vector, and tensor fluctuations of the metric.  It is
straightforward to
show that the scalar and vector pertubations do not contribute to
$R\tilde R$, and so we ignore these fluctuations in the following discussion.
We can also fix a gauge so that the tensor fluctuation is parameterized by the
two physical transverse traceless elements of $h_{ij}$.  For gravity
waves moving in the $z$ direction, we write
\bea\label{mytensors}
ds^2&=&-dt^2+a^2(t)\bigl[(1-\hp)dx^2\cr
&+&(1+\hp)dy^2 + 2\hc dxdy +dz^2\bigr]
\eea
where $a(t)  = e^{Ht}$ during inflation and $\hp$, $\hc$ are functions of
$t$, $z$.
To see the CP violation more explicitly, it is convenient to use a
helicity basis
\be
% h_L = {1\over \sqrt{2} } (\hp - i \hc) \ , \qquad
% h_R = {1\over \sqrt{2} } (\hp + i \hc) \ .
h_L =  (\hp - i \hc){/\sqrt{2}} \ , \qquad
h_R = (\hp + i \hc){/\sqrt{2}} \ .
\ee
Here $h_L$ and $h_R$ are complex conjugate scalar fields. To be very explicit,
the negative frequency part of $h_L$ is the conjugate of the positive
frequency part of $h_R$, and both are built from wavefunctions for left-handed
gravitons.

The contribution of tensor perturbations to $R\tilde R$, up to
second order in $h_L$ and $h_R$, is
\be\label{RRdual}
\begin{split}
%R\tilde R=& \frac{4i}{a^{3}}\biggl[
%\left(\frac{\partial^2}{\partial z^2}h_R
%\frac{\partial^2}{\partial t\partial z}h_L -
%\frac{\partial^2}{\partial z^2}h_L
%\frac{\partial^2}{\partial t\partial z}h_R\right)\cr
%&+ a^2\left(\frac{\partial^2}{\partial t^2}h_R
%\frac{\partial^2}{\partial t\partial z}h_L -
%\frac{\partial^2}{\partial t^2}h_L
%\frac{\partial^2}{\partial t\partial z}h_R\right)\cr
%&+\!\!\frac{1}{2}\frac{\partial}{\partial t}a^2\left(
%\frac{\partial}{\partial t}h_R\frac{\partial^2}{\partial t\partial z}
%h_L-\frac{\partial}{\partial t}h_L
%\frac{\partial^2}{\partial t\partial z}h_R\right)\biggr]
R\tilde R= \frac{4i}{a^{3}}\biggl[&\bigg(
{\partial^2_z} h_R\ {\partial_{z}\partial_{t}}h_L+a^2 {\partial^2_{t}}h_R\
{\partial_{t}\partial_{z}}h_L\cr
& +\frac{1}{2}{\partial_t}a^2 {\partial_t} h_R\
{\partial_{t}\partial_{z}}h_L\bigg)
-{\left(L\leftrightarrow R\right)}
\biggr]
\end{split}
\ee
If $h_L$ and $h_R$ have the same dispersion relation, this expression
vanishes. Thus, for $R\tilde R$  to
be nonzero, we need a
 `cosmological birefringence' during inflation.  Such an effect is induced
by the addition of \eqref{axioncoupling} to the gravitational
equations~\cite{LWK}.

Specifically, by adding \eqref{axioncoupling} to the Einstein action,
inserting \eqref{mytensors}, and varying with respect to the metric
fluctuations, we find the equations of motion
\be\label{LReqs}
  \ybox\, h_L = - 2i \frac{\Theta}{ a} {\dot h}^\prime_L \ , \qquad
  \ybox\, h_R = + 2i \frac{\Theta}{a} {\dot h}^\prime_R \ ,
\ee
where
\be\label{Thetaval}
  \Theta = 4 (F'' \dot\phi^2 + 2 H F'\dot\phi)/M_\Pl^2 \ ,
\ee
dots denote time derivatives, and primes denote differentiation of $F$
with respect to $\phi$.  Note that \eqref{axioncoupling} with a constant
$\phi$ is a total divergence that cannot affect the equations of motion;
thus, all terms in $\Theta$ involve derivatives of $\phi$.
 We have dropped terms with third-order derivatives
of $h_L$ and $h_R$ and terms with $\ddot \phi$. In fact, it is also
permissible to ignore the $F''$ term in \eqref{Thetaval}, since in
slow-roll inflation, $\dot \phi \ll M_\Pl H$ and each derivative on $F$
brings a dimensionful factor of order $1/M_\Pl$ \cite{F''}.
These  equations should be compared to those for evolution
in flat space
given by LWK~\cite{LWK}.
The new term proportional to $H\dot\phi$
leads to a substantial
enhancement in the size of $\Theta$.   With this
simplification, and the approximate form \eqref{Fval},
\be\label{myTheta}
\Theta = \sqrt{2\epsilon}{\cal N} (H/M_\Pl)^2/2\pi^2 \ ,
\ee
 where
$\epsilon = \half (\dot\phi)^2/(H M_\Pl)^2$
is the slow-roll parameter of inflation~\cite{LL}.

Let us now focus on the evolution of $h_L$ and, more specifically,
on its positive frequency component.
It is convenient to introduce conformal time
\be
%           \eta = {1\over H a}  = {1\over H} e^{-Ht} \ .
           \eta = {1/H a}  = {e^{-Ht}/H} \ .
\ee
(Note that conformal time $\eta$ runs in the opposite direction from $t$.)
The evolution equation for $h_L$ becomes
\be\label{hLformula}
  {d^2\over d\eta^2} h_L - 2 {1\over \eta} {d\over d\eta} h_L
- {d^2\over dz^2} h_L
          =  -2i\Theta {d^2\over d\eta dz} h_L \
\ee

If we ignore $\Theta$ for the moment and let $h_L \sim e^{ikz}$, this
becomes the equation of a spherical
Bessel function:
\be
  {d^2\over d\eta^2} h_L - 2 {1\over \eta} {d\over d\eta} h_L  + k^2 h_L = 0
\ee
for which the positive frequency solution is
\be\label{Bessel}
h_L^{+}(k,\eta) =  e^{+ i k(\eta+z)} (1 - i k\eta)  \ .
\ee

We now look for solutions  to \eqref{hLformula}
with $h_L \sim e^{ikz}$.  To do this, let
\be\label{gdef}
     h_L = e^{ikz} \cdot (-ik\eta)
                        e^{k\Theta\eta}  g(\eta)
\ee
Then $g(\eta)$ satisfies the equation
\be\label{Coulomb}
   {d^2\over d\eta^2} g + \left[ k^2 (1-\Theta^2) - {2\over \eta^2}
                 - { 2 k \Theta\over \eta} \right]\, g\ =\ 0 \ .
\ee
This is the equation of a Schr\"odinger particle with $\ell = 1$ in a
weak Coulomb potential.
%(the $Theta$ terms).
%When $\Theta = 0$, the Coulomb term vanishes and we
%find the spherical Bessel function \eqref{Bessel}.
For $h_L$, the Coulomb term is repulsive; for $h_R$, with the opposite sign
of the $\Theta$ term, the Coulomb potential is attractive.
This leads to attenuation of $h_L$ and amplification of $h_R$ in the early
universe. This is just the cosmological birefringence described by LWK~\cite{LWK}.

It will turn out that the generation of the matter asymmetry is dominated
by modes at short distances (sub-horizon modes) and at early times. This
corresponds to the limit $k\eta \gg 1$.
In this region, we can ignore the potential
terms in \eqref{Coulomb} and take the solution to be approximately
a plane wave.  More explicitly,
\be\label{findg}
    g(\eta) =   \exp[ ik(1-\Theta^2)^{1/2} \eta (1 + \alpha(\eta))] \ ,
\ee
where $\alpha(\eta) \sim \log \eta/\eta$.

%%\vspace*{1mm}
%\begin{center}
%{\bf{Green's function}}
%\end{center}
%\vspace*{1mm}
%\section{Green's function}

We would like to apply the results of \eqref{gdef} to compute
the expectation value of $R\tilde R$ in the inflationary space-time.
Our expression will be dominated by the quantum part of the gravity-wave
evolution.  For this regime, we can calculate the expectation value by
contracting $h_L$ and $h_R$ in $R\tilde R$ using an appropriate Green's
function. Define
\be\label{Gdefin}
\begin{split}
   G(x,t;x',t') &= \VEV{h_L(x,t) h_R(x',t')} \cr
                & = \int \, {d^3 k\over (2\pi)^3}
                 e^{ik \cdot (x-x')} G_k(\eta,\eta')\ .
\end{split}
\ee
For $k$ parallel to $z$, the Fourier component $G_k$ satisfies
\eqref{hLformula} with a delta-function source
\be\label{Gequation}
  \left[{d^2\over d\eta^2} - 2 ({1\over \eta}+k\Theta) {d\over d\eta}
+ k^2 \right]G_k(\eta,\eta')  =
          i  {(H\eta)^2\over M_\Pl^2} \delta(\eta - \eta') .
\ee
For $\Theta = 0$, the solution of this equation is
\be\label{Gvalue}
   G_{k0}(\eta, \eta') = \left\{\begin{array}{cc}
             (H^2/2k^3 M_\Pl^2) h_L^{+}(k,\eta) h_R^{-}(-k,\eta')
                   &   \quad \eta < \eta' \cr
               (H^2/2k^3 M_\Pl^2) h_L^{-}(k,\eta) h_R^{+}(-k,\eta')
                   &   \quad \eta' < \eta\ ,
\end{array}\right.
\ee
where $h_L^{-}$ is the complex conjugate of \eqref{Bessel}, and
$h_R^{+}$, $h_R^{-}$ are the corresponding solutions of the $h_R$ equation.
For $\Theta = 0$, these solutions are the same as for $h_L$, but the
structure of \eqref{Gvalue} will be preserved when we go to the case
$\Theta \neq 0 $.
The leading effect of $\Theta$ is to introduce
the exponential dependence from \eqref{gdef},
\be\label{myGk}
G_k=e^{-k\Theta\eta} G_{k0} e^{+k\Theta \eta'}
%      G_k = \left\{\begin{array}{cc}
%e^{-k\Theta\eta} G_{k0} e^{+k\Theta \eta'}
%                                &   \quad \eta < \eta' \cr
%                         e^{-k\Theta\eta} G_{k0} e^{+k\Theta \eta'}
%                                &   \quad\eta' < \eta
%\end{array}\right.
\ee
for both $\eta >\eta'$ and $\eta <\eta'$.
The prefactor is modified in order $\Theta^2$, and the wavefunctions
acquire additional corrections that are subleading for $k\eta \gg 1$.
Neither of these effects will be important for our result.

The Green's function \eqref{myGk}
can now be used to contract $h_L$ and $h_R$ to
evaluate the quantum expectation value of $R\tilde R$.  The result is \cite{Azadeh}
\be\label{RRdualval}
  \VEV{R\tilde R} =  {16\over a^4}\, \int \, {d^3 k\over (2\pi)^3}\
  {H^2\over 2 k^3 M_\Pl^2}
                      \cdot k^4 \Theta  + {\cal O}(\Theta^3)
\ee
where we pick up only the leading behavior for  $k\eta \gg 1$.

We note again that our expression for $\VEV{R\tilde R}$
is nonzero because of the effect of inflation in producing a CP
asymmetry out of equilibrium.  The
original
quantum state for the inflaton might have had nonzero amplitude for a
range of values of $\phi$ and might even have been CP-invariant.
However, inflation collapses the wavefunction onto a particular value of
$\phi$ that is caught up in the local expansion of the universe.  This
value gives us a classical background that is CP-asymmetric.

The above result and computations seems to be crucially depending on the form of the Green's function or the vacuum state we have used. To resolve the possible ambiguity in this regards, one may  perform the above computation using a different method, the fermion level crossing, e.g. following \cite{Gibbons-Steif}. This computation  confirms the above results \cite{to-appear}.

Inserting \eqref{RRdualval} into \eqref{Jlepton} and
integrating over the time period of inflation,
we find for the net lepton number density
\be\label{netlept}
  n = \int^{H^{-1}}_0 d\eta\ \int \, {d^3 k\over (2\pi)^3}\
       {3\over 16\pi^2}\, {16 H^2  k  \Theta \over M_\Pl^2}  \ .
\ee
The integral over $k$ runs over all of momentum space, up to a UV cutoff scale
at which our effective Lagrangian description breaks down. Let us denote this cutoff on physical momentum by $\mu$, i.e. $k\eta<\mu/H$.
The dominant effect comes not from the usual modes outside the horizon at the
end of inflation (super-horizon modes), $k\eta < 1$,
 but rather from very short distances compared
to these scales.  The integral over $\eta$ is dominated at large values
of $\eta$, early times.  The integral represents a compromise between two
effects of inflation, first, to blow up distances and thus carry us to
smaller physical momenta and, second, to dilute the generated lepton number
through expansion.   It is now clear that the dominant contribution to the
right-hand side comes from $1\ll k\eta <\mu/H$, as we had anticipated.  Performing
the integrals, we find
\be\label{nval}
      n  =  {3\over 48\pi^4} \left({ H\over
            M_\Pl}\right)^2 \Theta H^3  \left({\mu\over H} \right)^4 \ .
\ee
We might interpret this result physically in the following way.  The factor of 3 counts the mismatch between left and right handed degrees of freedom. The factor
$(H/M_\Pl)^2$ is the usual magnitude of the gravity wave power spectrum.
The factor $\Theta$ gives the magnitude of effective CP violation. The
factor $H^3$ is the inverse horizon size at inflation; this gives the density
$n$ appropriate units.  Finally, the factor $({\mu/H})^4$ gives the
enhancement over one's first guess due to our use of strongly quantum,
short distance fluctuations to generate $R\tilde R$, rather than the super-horizon modes
which effectively behave classically.

To understand the significance of this estimate, we should compare it to the
entropy density of the universe just after reheating, assuming that the
energy of the inflationary phase has been converted to the heat of a  gas of
massless particles. To estimate this, assume very naively that reheating
is instantaneous.  Then reheating converts an energy density
$\rho= 3 H^2 M_\Pl^2$ to radiation with
$\rho = \pi^2 g_* T^4/30$ and $s =  2\pi^2 g_* T^3/45$, where
$g_*$ is the effective
 number of massless degrees of freedom.  This gives
$ s = 2.3 g_*^{1/4} (H M_\Pl)^{3/2}$. With this value
\cite{Linde},
\be\label{novers}
    {n/ s} = 2.8\times 10^{-4} g_*^{-1/4} \left({H/ M_\Pl}\right)^{7/2}
                       \ \Theta \  \left({\mu/ H}\right)^4 \ .
\ee
%If we are less naive, we might follow the dilution of $n$ and $\rho$ with
%the expansion of the universe to the end of reheating.  The final result is
%the same.
Assuming that there have been no large increases in the entropy of the universe
since the end of reheating, \eqref{novers}
 can be compared directly to
the present value of $n/s$ inferred from \eqref{baryond}.  For this
one should note that the ratio of the
present baryon number to the lepton number originally generated in
leptogenesis is approximately $n_B/n = 4/11$ \cite{KRS}; then
\eqref{baryond} implies $n/s = 2.4 \times 10^{-10}$.

Inserting the estimate for $\Theta$ given in
\eqref{myTheta} and setting $g_* \sim 100$, we find
\be\label{mynovers}
    {n/ s } \sim  6.3 \times 10^{-6} \cdot \sqrt{\epsilon} {\cal N}
      \left({H/ M_\Pl}\right)^{11/2}  \left({\mu/ H}\right)^4 \ .
\ee
In principle, ${\cal N}$ can be a large dimensionless number, though
values greater than 1000 are probably difficult to accommodate within
string theory.  The ratio $(H/M_\Pl)$ is limited in simple slow-roll
inflation from the relation  $\delta \rho/\rho \sim  (H/M_\Pl)/\sqrt{\epsilon}
\sim 10^{-5}$.  The WMAP results give a more precise version of this bound
for the case of single-field inflation:
$H/M_\Pl < 1 \times 10^{-4}$~\cite{WMAPtwo}, implying that $\epsilon\sim 10^{-2}$ and hence
\be\label{novers-2nd}
{n/s } \sim  6.3 \times 10^{-4} \cdot \left({H/M_\Pl}\right)^{3/2}
\left({\mu/M_\Pl}\right)^4 \ .
\ee

Note that in \eqref{novers-2nd}, we assumed that the mass scale from the
kinetic term for the modulus
is the Planck or string scale.  If this mass scale is set at a lower
mass ${\cal F}$, $\Theta$
can be larger, scaling as
\be
    \Theta \sim {\sqrt{2\epsilon}{\cal N}}
         \left(H^2/{M_\Pl}{\cal F}\right)/{2\pi^2}
\sim \left({{H}/{M_\Pl}}\right)^2
\left({{M_\Pl}/{{\cal F}}}\right).
\ee
Assuming that ${\cal F}\sim \mu$ we obtain
\be\label{novers-3rd}
{n\over s } \sim  6.3 \times 10^{-4} \cdot \left({H\over M_\Pl}\right)^{3/2}
\left({\mu\over M_\Pl}\right)^3 \ .
\ee
Eq.~\eqref{novers-3rd} is our final result in which  $n/s \propto H^{3/2}$, corresponding to
$n\propto H^3$. As the first estimate, put
$\mu \sim H \sim 2\times 10^{14}$ GeV; this yields
$
n/s \sim 10^{-22} ,
$
a very small and unsatisfactory result.
Noting \eqref{novers-3rd}, this estimate can be improved by
taking higher values of $\mu$.  To recover the
observed value of $n/s= 2.4\times 10^{-10}$, $(\mu/ M_\Pl)^2\sim  10^{-5} (M_\Pl/H)$, assuming $H$ saturating its current upper bound,
$\mu\sim M_\Pl$. While smaller values of the cutoff $\mu$ are more favorable from particle physics viewpoint, this cutoff $\mu$ is within the acceptable range in string theory settings.

In summary, we have presented a new mechanism for the production of the
cosmic baryon asymmetry.  This mechanism relies on the axial vector anomaly
to violate fermion number and on the initial state of inflation to both for
CP violation and for out-of-equilibrium dynamics.   These are very
minimal ingredients that might be found in  a wide variety of models of
physics at short distances.
%The simple version of this model that we have analyzed here generates an  acceptable values
%for the matter asymmetry  only if we can make use of gravitational fluctuations
%somewhat above the usual grand unification scale.
It is interesting to
ask whether the conditions we have found can be embedded in a string theory motivated grand unification
model in a natural way.

\begin{acknowledgments}
We would like to thank
%our colleagues at SLAC and Stanford, particularly
N. Afshordi, E. Farhi, Y. Farzan, S. Kachru, L. McAllister, E. Silverstein,
S. Thomas, W. Unruh, R. Wagoner and especially R. Brandenbuger
for fruitful discussions.
We are grateful to R. Brandenberger and L. Smolin for their
encouragement during the course of this project.  We would like to especially thank A. Maleknejad for catching a mistake in the cutoff $\mu$ dependence of our final results \cite{Azadeh}.
The work of MMSh-J is supported in
part by the US NSF grant PHY--9870115 and by
funds from the Stanford Institute for Theoretical Physics.  The work
of SHSA and MEP is supported by the US DOE under
grant DE--AC03--76SF00515.
\end{acknowledgments}


\begin{thebibliography}{99}
\bibitem{F''}
Note that for non-linear $F(\phi)$, when the ``string scale'' ${\cal F}$ is much smaller than
$M_{pl}$, $F''$ may become important.
\bibitem{Linde}
If the reheating occurs slowly to a reheat temperature $T_r\ll T$,
then $n/s$ may be diluted with respect to this estimate by $T_r/T$. We
thank A. Linde for this comment.
\end{thebibliography}
\end{document}